\documentclass[aps,prb,onecolumn,superscriptaddress,showpacs,floatfix,amsmath,amssymb,reprint]{revtex4-1}
\usepackage{graphicx} 
\usepackage{bm}
\usepackage{color,mathtools,bbm} 
\usepackage{wasysym}	
\usepackage{hyperref}  
\usepackage{lineno}
\usepackage{hypernat}
\usepackage{float} 
\usepackage{wrapfig}   
\usepackage{mciteplus}

\usepackage{background}
\backgroundsetup{
  position=current page.east,
  angle=-90,
  nodeanchor=east,
  vshift=-5mm,
  opacity=0.5,
  scale=2,
  contents=Nanoscale -2019 DOI: 10.1039/C9NR03288F
}

\begin{document}


\title{Surface passivated and encapsulated ZnO atomic layer by high-$\kappa$ ultrathin MgO layer}

\author{C. E. Ekuma}
\email{che218@lehigh.edu}
\affiliation{Department of Physics, Lehigh University, Bethlehem, PA 18015}

\author{S. Najmaei}
\affiliation{Sensors and Electron Devices Directorate, United States Army Research Laboratory, Adelphi, MD 20783}

\author{M. Dubey}
\affiliation{Sensors and Electron Devices Directorate, United States Army Research Laboratory, Adelphi, MD 20783}

\date{\today}
\begin{abstract}  
\noindent 
Atomically transparent vertically aligned ZnO-based van der Waals material have been developed by surface passivation and encapsulation with atomic layers of MgO using materials by design; the physical properties investigated. The passivation and encapsulation led to a remarkable improvement in optical and electronic properties. The valence-band offset $\Delta E_v$ between MgO and ZnO, ZnO and MgO/ZnO, and ZnO and MgO/ZnO/MgO heterointerfaces are determined to be 0.37 $\pm$0.02, -0.05$\pm$0.02, and -0.11$\pm$0.02 eV, respectively; the conduction-band offset $\Delta E_c$ is deduced to be 0.97$\pm$0.02, 0.46$\pm$0.02, and 0.59$\pm$0.02 eV indicating straddling type-I in MgO and ZnO, and staggering type-II heterojunction band alignment in ZnO and the various heterostructures. The band-offsets and interfacial charge transfer are used to explain the origin of $n$-type conductivity in the superlattices. Enhanced optical absorption due to carrier confinement in the layers demonstrates that MgO is an excellent high-$\kappa$ dielectric gate oxide for encapsulating ZnO-based optoelectronic devices.
\end{abstract}  
    
\maketitle
Wide-bandgap semiconductors such as ZnO with a bulk direct bandgap of 3.37 eV are intensively being studied because they are promising materials for use in blue and ultraviolet light-emitting diodes (LEDs) and laser diodes~\cite{Huang1897,doi:10.1063/1.2977478}. The bandgap of ZnO can be tuned by substituting divalent element on the cation site. For example, alloying with Mg can increase the bandgap of bulk ZnO to as high as 4.0 eV whereas Cd alloying could decrease the bandgap to $\sim\,$3.0 eV~\cite{doi:10.1063/1.1992666,5325807}. To improve the device performance of LEDs and other ZnO-based devices, heterostructures have been fabricated and studied~\cite{doi:10.1063/1.2204655,doi:10.1063/1.3385384} but challenges such as reproducibility and low electroluminescence~\cite{Tsukazaki2004,doi:10.1002/adma.200502633} diminish scalability and efficiency. Device variability can be reduced by optical and electric confinement with a barrier material. MgO is an excellent passivating, high refractory, and a high-$\kappa$ dielectric material with notable existing technology base because it is a gate dielectric candidate for advanced complementary metal-oxide-semiconductor devices~\cite{Parkin2004,Yuasa2004}. 
Moreover, ZnMgO alloy can be lattice-matched with ZnO and a suitable barrier material to design ZnO/ZnMgO superlattices that enabled confinement of carriers and photons within the layers~\cite{doi:10.1063/1.1315340,Makino_2005}. Understanding the nature of the heterojunction of ZnO-based superlattices is essential in device applications. Perhaps, the most fundamental property of a heterojunction is the energy band-offsets, which for ZnO-based superlattices have been investigated by several experiments~\cite{doi:10.1063/1.2977478,doi:10.1063/1.2924279,doi:10.1063/1.1315340,Makino_2005}. Despite the need, there is no rigorous \textit{ab initio} investigation of the microstructural morphology, band-alignment and other critical parameters of the heterointerface like the nature of the charge transfer that could aid in understanding, e.g., the type of conductivity in undoped MgZnO alloy.

Herein, we report \textit{ab initio} investigation of the energy band-level alignment of ZnO-based van der Waals (vdW) heterostructure and the passivating characteristic of MgO in encapsulating ZnO. We provide a detailed analysis of the heterointerface and the band offsets of MgO and ZnO; ZnO and MgO/ZnO; ZnO and MgO/ZnO/MgO. Instead of using thin films or epilayers to construct the superlattices, we adopt single-atomic layers of 2D-based materials, which will provide natural quantum confinement that will reduce uncontrolled variability that may affect device performance. Additionally, based on our simulations, monolayer MgO (the passivating layer) exhibit indirect and wide bandgap; this can mitigate against parasitic absorption losses and unwanted backscattering processes, which competes with band-to-band absorption to decrease photocurrent.~\cite{doi:10.1063/1.4861404}.      

The ZnO-based systems were grown by vertical aligning one-atomic layers of ZnO and MgO to form vdW heterostructures of MgO/ZnO and MgO/ZnO/MgO, respectively. The electronic properties and the interfacial characteristics are studied using \textit{first-principles} based on density functional theory~\cite{Hohenberg1964,*Kohn1965} as implemented in {\it Vienna Ab Initio Simulation Package}~\cite{KRESSE199615} using HSE06 hybrid functional~\cite{doi:10.1063/1.2404663,*doi:10.1063/1.2204597}, a kinetic energy cutoff of 550\,eV for the plane-wave basis set, and a $4\times 4\times 1$ Monkhorst-Pack grid to represent the reciprocal space. Structural relaxations were carried out using the Perdew-Burke-Ernzerhof~\cite{PhysRevLett.77.3865} exchange-correlation functional, and atomic positions are allowed to relax until the energy (charge) is converged to within $\sim 10^{-3}$ ($10^{-7}$) eV and the forces dropped to $\sim 10^{-3}$ eV/\,{\AA}. All calculations included van der Waals interactions, dipole corrections, and a vacuum of $ 25$\,{\AA} along the out-of-plane direction to eliminate the artifacts of the periodic boundary condition.

\begin{figure*}
 \includegraphics[trim = 0mm 0mm 0mm 0mm,scale=0.5,keepaspectratio,clip=true]{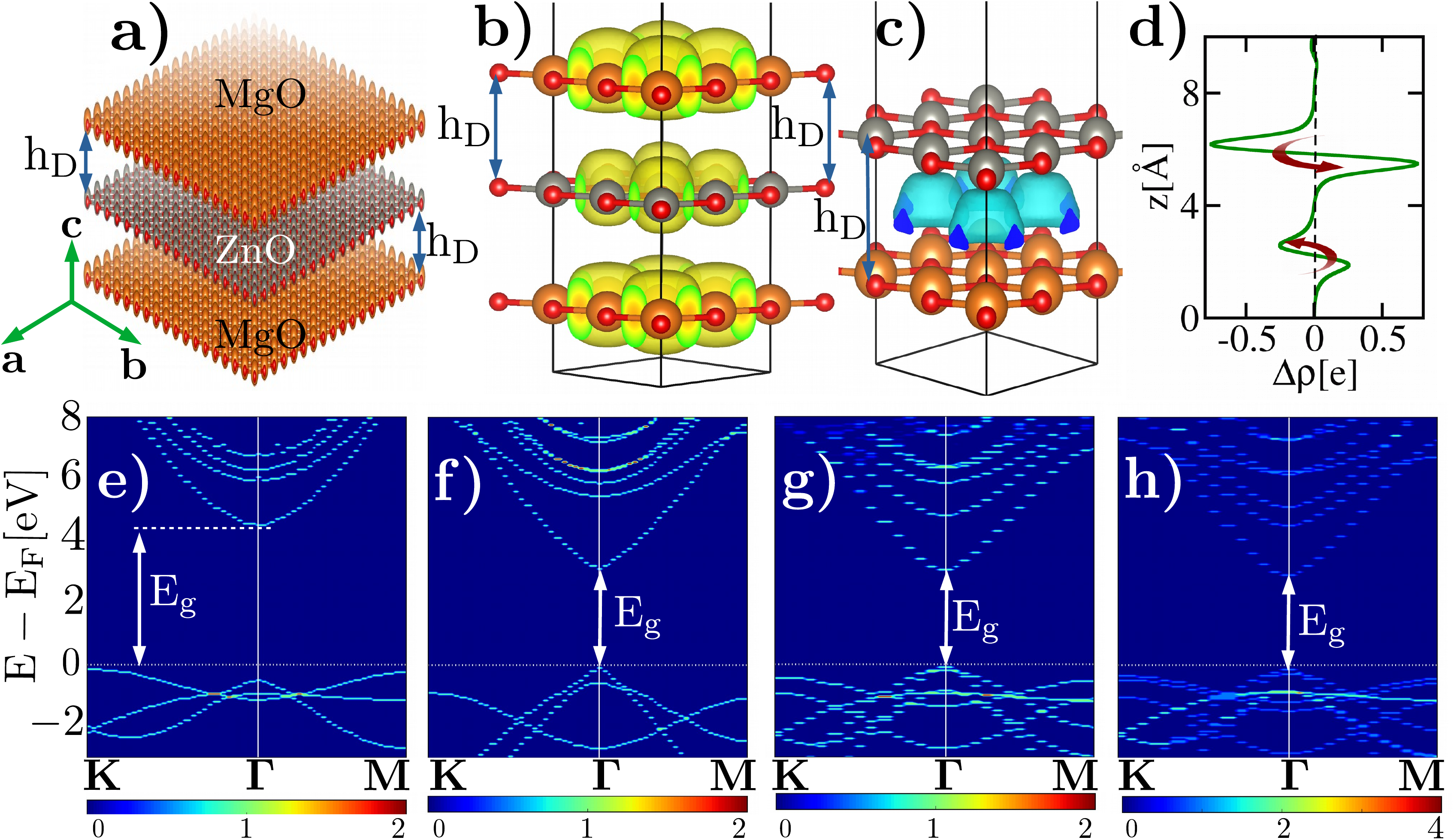}
\caption{\textbf{Crystal lattice and electronic spectra.}  Crystal lattice of MgO encapsulated ZnO without \textbf{(a)} and with \textbf{(b)} the ELF of MgO/ZnO/MgO vdW heterostructure; \textbf{(c)} is the difference of ELF for MgO/ZnO vdW heterostructure showing the formation of charge depletion/accumulation layer. The yellow (cyan) lobes indicate the positive (negative) phase with an arbitrary isosurface value. \textbf{(d)} Planar-integrated charge density difference $\Delta\rho(z)$ defined as the difference between the charge densities of MgO/ZnO/MgO and the superposition of the isolated monolayer ZnO and MgO layers in unit of elementary charge $e$. \textbf{(e-h)} are the band structure of monolayer MgO, ZnO, ZnO/MgO, and MgO/ZnO/MgO, respectively. \textbf{g} and \textbf{h} have been unfolded into the first Brillouin of ZnO. In the plots, $\mathrm{h_D}$ is the interlayer separation calculated to be 3.58$\pm0.2\,${\AA}, $\mathrm{E_g}$ is the energy bandgap as shown in Table~\ref{tab} and Figure~\ref{f2}\textbf{a}. The arrows in Figure~\ref{f1}\textbf{(d)} is the most probable direction of the charge transfer.}
\label{f1}
\end{figure*}

We present in Figure~\ref{f1}\textbf{a} the optimized crystal structure for MgO/ZnO/MgO vdW heterostructure while the corresponding electron localization function (ELF) is presented in Figure~\ref{f1}\textbf{b}. The difference of ELF for MgO/ZnO vdW heterostructure is shown in Figure~\ref{f1}\textbf{c}. The vdW heterostructures are designed from a commensurate $2\times 2\times 1$ supercell slab using optimized crystal structure of monolayer ZnO and MgO, respectively, with only $\sim0.49\,$\% lattice mismatch. Our simulation show absence of buckling or distortion in either layer and no chemical bonding at the interface (Figure~\ref{f1}\textbf{b}). The interlayer separation is calculated to be 3.58$\pm\,$0.20 {\AA}, which confirms an interface dominated by proximity effects through van der Waals interactions. We find the charges are localized at each layer (Figure~\ref{f1}\textbf{b}) with depletion/accumulation of charges across the heterojunction (Figure~\ref{f1}\textbf{c}). This observation is supported by the highly nonlinear charge redistribution in the planar-integrated charge density difference $\Delta\rho(z)$ of the superlattice and the constituent layers, where $\rho(z) =\iint \rho(x,y,z)dx\,dy$ (Figure~\ref{f1}\textbf{d}). 

To understand the electronic structure, we show in Figure~\ref{f1}\textbf{e}\&\textbf{f} the calculated electronic band structure of monolayer MgO and ZnO while that of the MgO/ZnO and MgO/ZnO/MgO vdW heterostructures unfolded unto the Brillouin zone (BZ) of monolayer ZnO are presented in Figure~\ref{f1}\textbf{g}\&\textbf{h}. The predicated bandgaps are listed in Table~\ref{tab}. We observed nontrivial renormalization of the electronic structure of the vdW heterostructures including the decrease of the energy bandgap $E_g$, which remained direct at the $\Gamma$-point of the first BZ. As in conventional ZnO-based heterostructures, one would expect the E$_g$ to increase due to Mg doping~\cite{doi:10.1063/1.1992666,5325807}. However, we note the markedly different interlayer bonding interactions at play in conventional and van der Waals heterostructures; strong covalent or ionic versus weak van der Waals interactions in conventional and layered 2D-based heterostructures, respectively. To gain an understanding of the low-energy states, we characterize the contributions of each of the atoms around the Fermi level. The states in the proximity of the Fermi level are dominated by a strong hybridization of Zn-($s$,$p$ \& $d$) states and O-$p$ states from the ZnO layer. The only significant contribution from the MgO layer in the proximity of the Fermi level is the Mg-($s$ \& $p$) states observed around 0.50-3.50 eV in the valence band [see Supplemental Materials (SM)]~\cite{sm}; evident by the increased scattering and blurring in the valence states (Figure~\ref{f1}\textbf{g\&h}). Even with the observed nontrivial effects in the electronic structure, the predicted effective masses of the vdW heterostructures are practically unchanged when compared to that of free-standing monolayer ZnO (Table~\ref{tab}), which is a sign that monolayer MgO will be a good passivating layer for ZnO-based nanoscale devices.

\begin{table}
\caption{Energy bandgap E$_g$, work function $\Phi$, and the built-in voltage $\Delta\Phi$ all in eV. Also shown is the effective electron $m_e$ and hole $m_h$ masses in units of free electron mass $m_o$. For clarity, we denote MgO/ZnO and MgO/ZnO/MgO vdW heterostructures as {\bf A} and {\bf B}, respectively.~\footnote{The work function is obtained as $\Phi=\mathrm{E_F}-\mathrm{E_{vac}}$, where $\mathrm{E_F}$ and $\mathrm{E_{vac}}$ is the Fermi-level and vacuum energy, respectively; the band effective masses $m^{e/h}$ are obtained by fitting a parabola to the band extremum of states around the Fermi level of Figure~\ref{f1}\textbf{e-h}.}}
\begin{tabular}{ccccccccc}
\hline \hline
Name & E$_g$ & $\Phi$ & $\Delta\Phi$ & $m_e$ & $m_h$ &  \\ 
\hline
ZnO & 3.58 & 5.90 &0.0 &0.34/0.34 & 1.03/0.99&  \\ 
MgO & 4.91 & 6.20 & 0.0 & 0.58/0.61 & 6.73/4.92&  \\ 
{\bf A} & 3.17 & 5.89 & 1.97 &0.33/0.33& 1.04/1.04&  \\ 
{\bf B} & 3.10 & 5.99 & 2.19 &0.31/0.31 & 1.04/1.05&  \\ 
\hline \hline
\end{tabular}
\label{tab}
\end{table}

Even though there is no obvious chemical bonding at the heterojunction, there is a significant charge transfer (CT) and charge redistribution due to proximity effects (Figure~\ref{f1}\textbf{c}\&\textbf{d}). A charge transfer calculations will provide a clearer picture of the carriers at the interface, which we obtained using the density derived electrostatic and chemical net atomic charges~\cite{C6RA04656H,*C6RA05507A}. In the MgO/ZnO vdW heterostructure, we predict a CT$\sim0.033\,|e|$ while for MgO/ZnO/MgO superlattice, we obtain an average CT$\sim0.038\,|e|$. In both superlattices, the electrons are transferred from the MgO to ZnO layer leading to $n$-type conductivity with an average doping concentration of $9.43\pm0.52\times10^{12}\,$e/cm$^2$. The direction of the charge transfer seems to be a contradiction since naively, one would assume the ZnO layer is doped with holes since the work function $\Phi$ of monolayer ZnO is smaller than that of MgO (Table~\ref{tab}). However, we note, the electron transfer from the MgO layer to that of ZnO is consistent with the fact that the electron affinity of MgO is smaller than the work function of ZnO, and would produce dipole in this direction. Our calculations further show that the charge transfer across the heterojunction needs to overcome an energy barrier of $5.90\pm0.02\,$eV and a transfer distance of $3.72\pm0.02\,${\AA} for MgO/ZnO. For MgO/ZnO/MgO, the charge transfer across the heterointerface needs to overcome an energy barrier of $4.84\pm0.02$ ($5.99\pm0.02\,$eV) and a transfer distance of $3.59\pm0.02$ ($3.72\pm0.02\,${\AA}) for the top (bottom) interfaces (see SM~\cite{sm}). This implies the carrier transfer across the heterojunction is energetically and spatially favorable. Remarkably, as observed from the electronic structure analysis, only small fraction of Mg-states from the MgO layer contribute to the states around the Fermi level leading to electron doping of the ZnO layer; this is reminiscent of low Mg content with $n$-type conduction experimentally observed in undoped MgZnO~\cite{doi:10.1063/1.2816914,*doi:10.1063/1.3485058}.

\begin{figure*}[bth]
 \includegraphics[trim = 0mm 0mm 0mm 0mm,scale=0.5,keepaspectratio,clip=true]{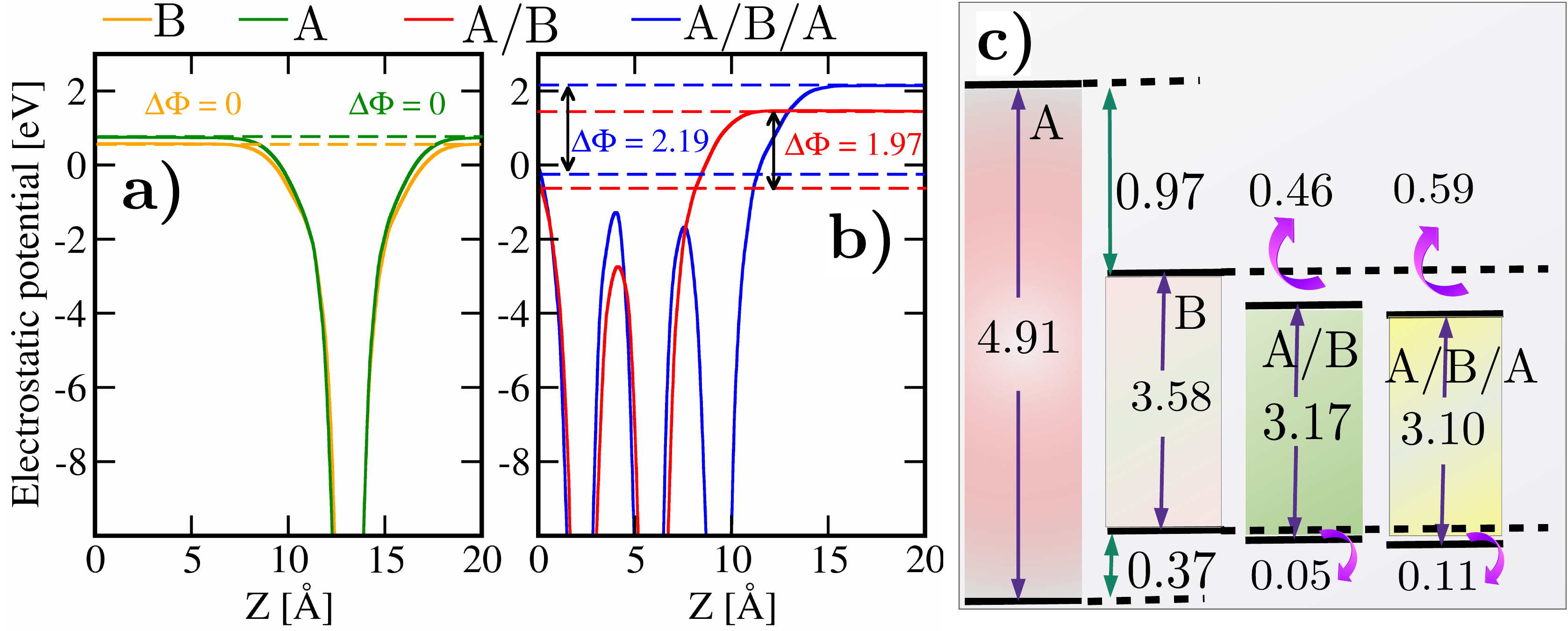}
\caption{\textbf{Electronic properties of ZnO-based vdW heterostructures.} Electrostatic potential profile along the $z$-direction of the crystallographic plane for \textbf{(a)} monolayer ZnO and MgO, and \textbf{(b)} MgO/ZnO and MgO/ZnO/MgO. Observe the absence (emergence) of built-in voltage $\Delta\Phi$ in the monolayers (heterostructures) due to the difference in the electronegativity of Zn and Mg atoms. \textbf{(c)} Energy band-level alignment diagrams; we predict straddling type-I heterojunction for MgO and ZnO, and staggered type-II heterojunction interfaces between ZnO and the vdW heterostructures of MgO/ZnO and MgO/ZnO/MgO. \textbf{A} and \textbf{B} denote MgO and ZnO. All values are in eV.}
\label{f2}
\end{figure*}

Band bending is associated with the density and energy distribution of the interfacial charges. The spatial charge redistribution due to charge transfer across the interface will also modify the work function with an accompanying potential step $\Delta\Phi$ (Figure~\ref{f2}\textbf{b}) attributed to the spontaneous polarization effect in the ZnO layer. Significant built-in voltage has been reported in ZnO-based superlattices~\cite{PhysRevB.72.241305}. Note that $\Delta\Phi$ is absent in the monolayers (c.f. Figure~\ref{f2}\textbf{a-b}). Before now, we are not aware of any data on the interfacial properties of atomically thin ZnO-based heterostructures. However, just like most 2D-based heterostructures, there is evidence of weak or absence of chemical interaction between the layers; an intrinsic interfacial dipole (Figure~\ref{f2}\textbf{a-b}) that lowers the vacuum level of monolayer MgO to that of monolayer ZnO and precisely compensates for the difference between their work function and the electron barrier. 

To understand the band-offset of the various structures, which is an essential property of an interface in device applications, we present in Figure~\ref{f2}\textbf{c} the energy band-level alignment diagrams, with all of the energy scales included. Between monolayer MgO and ZnO, the conduction band offset $\Delta E_c$ is determined to be 0.97$\pm0.02$ eV while the valence-band offset $\Delta E_v$ is estimated as 0.37 $\pm$0.02 eV, indicating a type-I alignment for MgO and ZnO heterojunction. The ratio 
$\Delta E_c:\Delta E_v$ is estimated as 5:2, which is smaller than $\sim$ 4:1 reported by experiment~\cite{doi:10.1063/1.2924279}. The difference could be attributed to many factors, e.g., (i) markedly different interlayer bonding in our 2D-based crystals as compared to the conventional crystals; and (ii) the MgO in the experimental has a cubic structure, but the MgO in our study has a hexagonal structure. We also obtained the band-levels alignment of MgO with the superlattices to be a straddling type-I heterointerface (Figure~\ref{f2}\textbf{c}). A good dielectric gate oxide for passivating ZnO-based devices should exhibit negligible hysteresis due to bias stressing~\cite{doi:10.1063/1.2962985,*doi:10.1063/1.2966145,5353711,*doi:10.1889/JSID18.10.753} with conduction and valence band offsets that is large enough to mitigate carrier injections into the dielectric~\cite{Robertson2000}. Our predicted type-I band-level alignments of MgO and ZnO including the superlattices meet the above criteria; demonstrating that MgO is a good passivating and  high-$\kappa$ oxide gate dielectric layer for ZnO-based nanoscale devices.

The straddling type-I band-level alignment between MgO and ZnO also shows that MgO has a higher $E_c$ and lower $E_V$ to ZnO; intuitively, a MgZnO alloy will exhibit lower $E_v$ and higher $E_c$, which surprisingly agrees with the band-level alignment we obtained for MgO/ZnO vdW heterostructure. Such a band-level arrangement has been attributed to lead to $p$-type conductivity in undoped MgZnO~\cite{doi:10.1063/1.2924279}. Our calculations reveal that the MgO/ZnO vdW heterostructure exhibits $n$-type conductivity; suggesting that the origin of $p$-type conductivity in undoped MgZnO alloy may be more complex such as Zn vacancies or higher Mg content.

\begin{figure}
 \includegraphics[trim = 0mm 0mm 0mm 0mm,scale=0.5,keepaspectratio,clip=true]{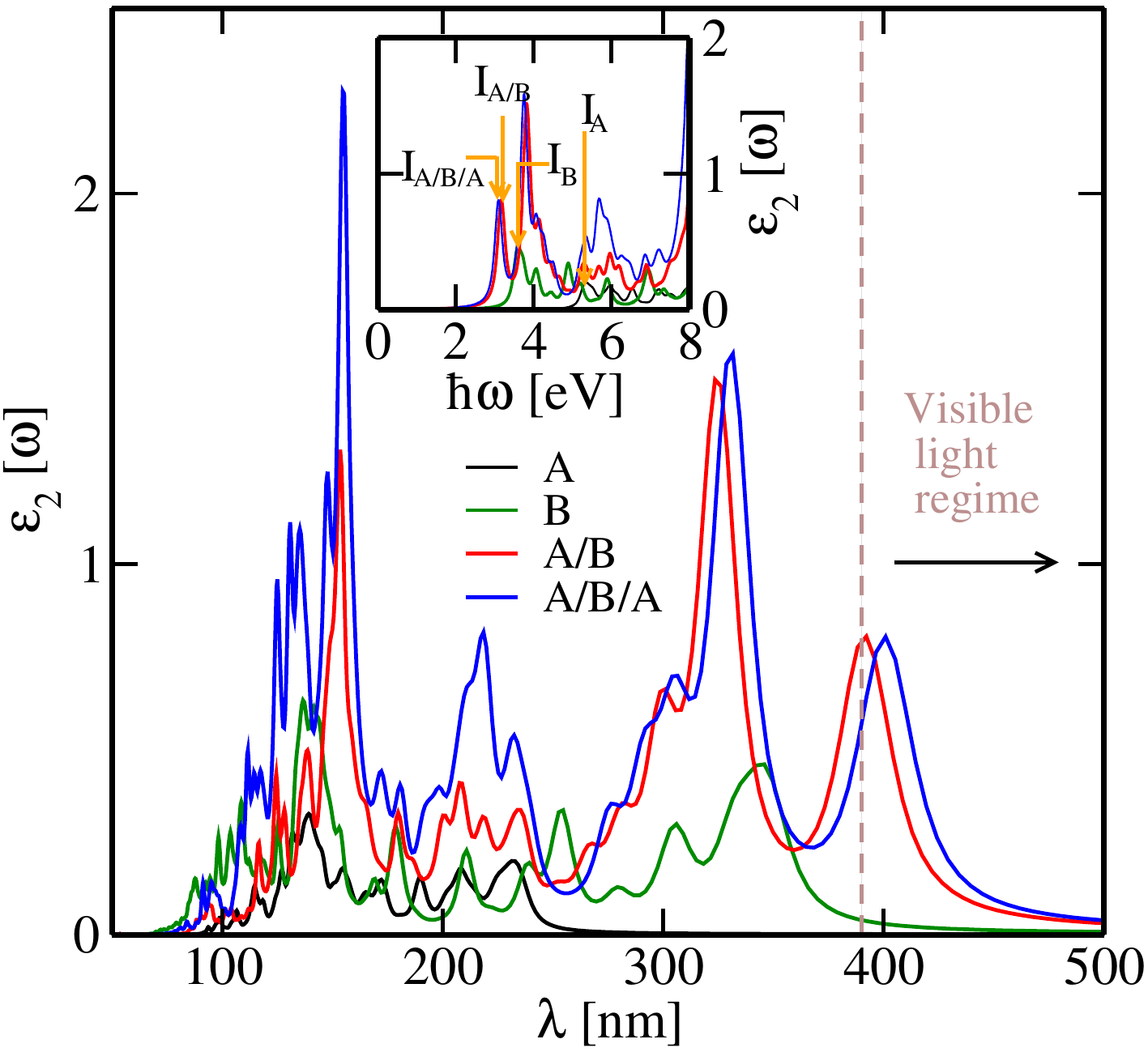}
\caption{\textbf{Optical properties of ZnO-based vdW heterostructures.} Absorptive part of the dynamical complex dielectric function $\varepsilon_2(\omega)$ versus photon wavelength $\lambda (nm)$ for MgO (A), ZnO (B), MgO passivated ZnO (A/B), and MgO encapsulated ZnO (A/B/A). The inset is $\varepsilon_2(\omega)$ versus photon energy $\hbar\omega$ showing the onset of the optical excitation estimated as $\mathrm{I_A}\sim5.32$, $\mathrm{I_B}\sim3.51$, $\mathrm{I_{A/B}}\sim3.15$, and $\mathrm{I_{A/B/A}}\sim3.06\,$eV, which is a measure of the optical bandgap. A and B denote MgO and ZnO.}
\label{f3}
\end{figure}

Next, we focus on the energy band alignment of the various superlattices. For MgO/ZnO vdW heterostructure, the bandgaps of monolayer ZnO and MgO/ZnO are 3.58 and 3.17 eV, respectively. So the $\Delta E_c$ and $\Delta E_v$ is estimated to be 0.46$\pm$0.02 and -0.05$\pm$0.02 eV, respectively, indicating that a type-II alignment for ZnO and MgO/ZnO heterointerface. The bandgap difference $\Delta E_g$ of 0.41 eV has an almost 9:1 ratio between $\Delta E_c$ and $\Delta E_v$ in basic agreement with experiment for ZnO and undoped low-Mg content of ZnOMgO superlattices~\cite{doi:10.1063/1.2816914,*doi:10.1063/1.3485058,doi:10.1063/1.124573}; but significantly higher than $\sim$7:3 reported for ZnO/MgZnO quantum well heterostructures~\cite{doi:10.1063/1.1370116,doi:10.1063/1.2977478}. For the MgO encapsulated ZnO superlattice, we also obtained a type-II heterointerface alignment with ZnO. The $\Delta E_c$ and $\Delta E_v$ is estimated to be 0.59$\pm$0.02 and -0.11$\pm$0.02 eV, respectively; and with $\Delta E_g\sim0.48\,$eV, we deduce $\approx6:1$ ratio between $\Delta E_c$ and $\Delta E_v$ of ZnO and MgO/ZnO/MgO. The determination of the band-level alignments indicates that ZnO and the superlattices are satisfactory for confining both electron and hole that is important for the design and application of ZnO-based nanoscale devices. We note that the heterojunction data of MgO encapsulated ZnO is not yet available as such, our predictions are crucial for experimental design and characterizations of nanoscale ZnO-based heterostructures for optoelectronic device applications.   

To gain a better understanding of the optoelectronic properties of the designed vdW materials, we calculated their absorption spectra. The optical absorption characterized by the absorptive part of the complex dynamical dielectric function $\varepsilon_2(\omega)$ versus wavelength reveals that after surface passivation and encapsulation, the light absorptivity of ZnO nanostructure is significantly increased (Figure~\ref{f3}). Upon passivation and encapsulation, the optical bandgap of the superlattices decreased; we estimate the optical bandgap of the pristine materials and the superlattices using the lowest sharp structure around their spectra corresponding to 5.32, 3.51, 3.15, and 3.06 eV for MgO, ZnO, MgO/ZnO, and MgO/ZnO/MgO, respectively, which show an overall blueshift (inset of Figure~\ref{f3}). Most significant, the passivated and encapsulated structures show significantly increased absorptivity that transcended well into the visible regime. We attribute the increased absorption to better confinement of carriers by the MgO barrier layer. 

In summary, we have designed and studied atomically thin layer vertically aligned ZnO-based van der Waals materials. Using \textit{ab initio} modeling, we predict straddling type-I heterojunction with a positive valence band offsets between monolayer MgO and ZnO; MgO and MgO/ZnO; and MgO and MgO/ZnO/MgO, which is a signature of a barrier for hole injection. Through a combination of charge transfer and band-levels diagram analyses, we demonstrate that MgO is a candidate for a) high-$\kappa$ oxide gate dielectric in ZnO-based nanoscale devices, and b) an excellent passivating barrier layer to encapsulate ZnO, which will eliminate surface-related device instabilities and aid carrier confinement.

\section*{Acknowledgement}
This work was supported by the Lehigh University Start-up fund to CEE and the U.S. Army Research Laboratory. Supercomputer support is provided by the DOD High-Performance Computing Modernization Program at the Army Engineer Research and Development Center, Vicksburg, MS.

\scriptsize{

 }

\begin{mcitethebibliography}{10}

\bibitem{Huang1897}
M.~H. Huang, S.~Mao, H.~Feick, H.~Yan, Y.~Wu, H.~Kind, E.~Weber, R.~Russo, and
  P.~Yang, Science {\bf 292}(5523), 1897--1899 (2001)\relax
\mciteBstWouldAddEndPuncttrue
\mciteSetBstMidEndSepPunct{\mcitedefaultmidpunct}
{\mcitedefaultendpunct}{\mcitedefaultseppunct}\relax
\EndOfBibitem
\bibitem{doi:10.1063/1.2977478}
S.~C. Su, Y.~M. Lu, Z.~Z. Zhang, C.~X. Shan, B.~H. Li, D.~Z. Shen, B.~Yao,
  J.~Y. Zhang, D.~X. Zhao, and X.~W. Fan, Applied Physics Letters {\bf 93}(8),
  082108 (2008)\relax
\mciteBstWouldAddEndPuncttrue
\mciteSetBstMidEndSepPunct{\mcitedefaultmidpunct}
{\mcitedefaultendpunct}{\mcitedefaultseppunct}\relax
\EndOfBibitem
\bibitem{doi:10.1063/1.1992666}
U.~\"{O}zg\"{u}r, Y.~I. Alivov, C.~Liu, A.~Teke, M.~A. Reshchikov,
  S.~Do\v{g}an, V.~Avrutin, S.-J. Cho, and H.~Morko\c{c}, Journal of Applied
  Physics {\bf 98}(4), 041301 (2005)\relax
\mciteBstWouldAddEndPuncttrue
\mciteSetBstMidEndSepPunct{\mcitedefaultmidpunct}
{\mcitedefaultendpunct}{\mcitedefaultseppunct}\relax
\EndOfBibitem
\bibitem{5325807}
Y.~{Choi}, J.~{Kang}, D.~{Hwang}, and S.~{Park}, IEEE Transactions on Electron
  Devices {\bf 57}(1), 26--41 (2010)\relax
\mciteBstWouldAddEndPuncttrue
\mciteSetBstMidEndSepPunct{\mcitedefaultmidpunct}
{\mcitedefaultendpunct}{\mcitedefaultseppunct}\relax
\EndOfBibitem
\bibitem{doi:10.1063/1.2204655}
M.-C. Jeong, B.-Y. Oh, M.-H. Ham, and J.-M. Myoung, Applied Physics Letters
  {\bf 88}(20), 202105 (2006)\relax
\mciteBstWouldAddEndPuncttrue
\mciteSetBstMidEndSepPunct{\mcitedefaultmidpunct}
{\mcitedefaultendpunct}{\mcitedefaultseppunct}\relax
\EndOfBibitem
\bibitem{doi:10.1063/1.3385384}
J.~B. You, X.~W. Zhang, S.~G. Zhang, H.~R. Tan, J.~Ying, Z.~G. Yin, Q.~S. Zhu,
  and P.~K. Chu, Journal of Applied Physics {\bf 107}(8), 083701 (2010)\relax
\mciteBstWouldAddEndPuncttrue
\mciteSetBstMidEndSepPunct{\mcitedefaultmidpunct}
{\mcitedefaultendpunct}{\mcitedefaultseppunct}\relax
\EndOfBibitem
\bibitem{Tsukazaki2004}
A.~Tsukazaki, A.~Ohtomo, T.~Onuma, M.~Ohtani, T.~Makino, M.~Sumiya, K.~Ohtani,
  S.~F. Chichibu, S.~Fuke, Y.~Segawa, H.~Ohno, H.~Koinuma, and M.~Kawasaki,
  Nature Materials {\bf 4}, 42 (2004)\relax
\mciteBstWouldAddEndPuncttrue
\mciteSetBstMidEndSepPunct{\mcitedefaultmidpunct}
{\mcitedefaultendpunct}{\mcitedefaultseppunct}\relax
\EndOfBibitem
\bibitem{doi:10.1002/adma.200502633}
J.-H. Lim, C.-K. Kang, K.-K. Kim, I.-K. Park, D.-K. Hwang, and S.-J. Park,
  Advanced Materials {\bf 18}(20), 2720--2724 (2006)\relax
\mciteBstWouldAddEndPuncttrue
\mciteSetBstMidEndSepPunct{\mcitedefaultmidpunct}
{\mcitedefaultendpunct}{\mcitedefaultseppunct}\relax
\EndOfBibitem
\bibitem{Parkin2004}
S.~S.~P. Parkin, C.~Kaiser, A.~Panchula, P.~M. Rice, B.~Hughes, M.~Samant, and
  S.-H. Yang, Nature Materials {\bf 3}, 862 (2004)\relax
\mciteBstWouldAddEndPuncttrue
\mciteSetBstMidEndSepPunct{\mcitedefaultmidpunct}
{\mcitedefaultendpunct}{\mcitedefaultseppunct}\relax
\EndOfBibitem
\bibitem{Yuasa2004}
S.~Yuasa, T.~Nagahama, A.~Fukushima, Y.~Suzuki, and K.~Ando, Nature Materials
  {\bf 3}, 868 (2004)\relax
\mciteBstWouldAddEndPuncttrue
\mciteSetBstMidEndSepPunct{\mcitedefaultmidpunct}
{\mcitedefaultendpunct}{\mcitedefaultseppunct}\relax
\EndOfBibitem
\bibitem{doi:10.1063/1.1315340}
A.~Ohtomo, K.~Tamura, M.~Kawasaki, T.~Makino, Y.~Segawa, Z.~K. Tang, G.~K.~L.
  Wong, Y.~Matsumoto, and H.~Koinuma, Applied Physics Letters {\bf 77}(14),
  2204--2206 (2000)\relax
\mciteBstWouldAddEndPuncttrue
\mciteSetBstMidEndSepPunct{\mcitedefaultmidpunct}
{\mcitedefaultendpunct}{\mcitedefaultseppunct}\relax
\EndOfBibitem
\bibitem{Makino_2005}
T.~Makino, Y.~Segawa, M.~Kawasaki, and H.~Koinuma, Semiconductor Science and
  Technology {\bf 20}(4), S78--S91 (2005)\relax
\mciteBstWouldAddEndPuncttrue
\mciteSetBstMidEndSepPunct{\mcitedefaultmidpunct}
{\mcitedefaultendpunct}{\mcitedefaultseppunct}\relax
\EndOfBibitem
\bibitem{doi:10.1063/1.2924279}
Y.~F. Li, B.~Yao, Y.~M. Lu, B.~H. Li, Y.~Q. Gai, C.~X. Cong, Z.~Z. Zhang, D.~X.
  Zhao, J.~Y. Zhang, D.~Z. Shen, and X.~W. Fan, Applied Physics Letters {\bf
  92}(19), 192116 (2008)\relax
\mciteBstWouldAddEndPuncttrue
\mciteSetBstMidEndSepPunct{\mcitedefaultmidpunct}
{\mcitedefaultendpunct}{\mcitedefaultseppunct}\relax
\EndOfBibitem
\bibitem{doi:10.1063/1.4861404}
J.~Peter~Seif, A.~Descoeudres, M.~Filipi\u{c}, F.~Smole, M.~Topi\u{c},
  Z.~Charles~Holman, S.~De~Wolf, and C.~Ballif, Journal of Applied Physics {\bf
  115}(2), 024502 (2014)\relax
\mciteBstWouldAddEndPuncttrue
\mciteSetBstMidEndSepPunct{\mcitedefaultmidpunct}
{\mcitedefaultendpunct}{\mcitedefaultseppunct}\relax
\EndOfBibitem
\bibitem{Hohenberg1964}
P.~Hohenberg and W.~Kohn, Phys. Rev. {\bf 136}, B864--B871 (1964)\relax
\mciteBstWouldAddEndPuncttrue
\mciteSetBstMidEndSepPunct{\mcitedefaultmidpunct}
{\mcitedefaultendpunct}{\mcitedefaultseppunct}\relax
\EndOfBibitem
\bibitem{Kohn1965}
W.~Kohn and L.~J. Sham, Phys. Rev. {\bf 140}, A1133--A1138 (1965)\relax
\mciteBstWouldAddEndPuncttrue
\mciteSetBstMidEndSepPunct{\mcitedefaultmidpunct}
{\mcitedefaultendpunct}{\mcitedefaultseppunct}\relax
\EndOfBibitem
\bibitem{KRESSE199615}
G.~Kresse and J.~Furthm\"{u}ller, Comput. Mater. Sci. {\bf 6}(1), 15--50
  (1996)\relax
\mciteBstWouldAddEndPuncttrue
\mciteSetBstMidEndSepPunct{\mcitedefaultmidpunct}
{\mcitedefaultendpunct}{\mcitedefaultseppunct}\relax
\EndOfBibitem
\bibitem{doi:10.1063/1.2404663}
A.~V. Krukau, O.~A. Vydrov, A.~F. Izmaylov, and G.~E. Scuseria, The Journal of
  Chemical Physics {\bf 125}(22), 224106 (2006)\relax
\mciteBstWouldAddEndPuncttrue
\mciteSetBstMidEndSepPunct{\mcitedefaultmidpunct}
{\mcitedefaultendpunct}{\mcitedefaultseppunct}\relax
\EndOfBibitem
\bibitem{doi:10.1063/1.2204597}
J.~Heyd, G.~E. Scuseria, and M.~Ernzerhof, The Journal of Chemical Physics {\bf
  124}(21), 219906 (2006)\relax
\mciteBstWouldAddEndPuncttrue
\mciteSetBstMidEndSepPunct{\mcitedefaultmidpunct}
{\mcitedefaultendpunct}{\mcitedefaultseppunct}\relax
\EndOfBibitem
\bibitem{PhysRevLett.77.3865}
J.~P. Perdew, K.~Burke, and M.~Ernzerhof, Phys. Rev. Lett. {\bf 77}, 3865--3868
  (1996)\relax
\mciteBstWouldAddEndPuncttrue
\mciteSetBstMidEndSepPunct{\mcitedefaultmidpunct}
{\mcitedefaultendpunct}{\mcitedefaultseppunct}\relax
\EndOfBibitem
\bibitem{sm}
See Supplemental Material at [URL will be inserted by publisher]. The partial density of states; the electronic band-level diagram depicting the charge transfer distance and energy barrier; and the lattice parameters in CIF format of the van der Waals materials.\relax
\mciteBstWouldAddEndPuncttrue
\mciteSetBstMidEndSepPunct
{\mcitedefaultendpunct}{\mcitedefaultseppunct}\relax
\EndOfBibitem
\bibitem{C6RA04656H}
T.~A. Manz and N.~G. Limas, RSC Adv. {\bf 6}, 47771--47801 (2016)\relax
\mciteBstWouldAddEndPuncttrue
\mciteSetBstMidEndSepPunct{\mcitedefaultmidpunct}
{\mcitedefaultendpunct}{\mcitedefaultseppunct}\relax
\EndOfBibitem
\bibitem{C6RA05507A}
N.~G. Limas and T.~A. Manz, RSC Adv. {\bf 6}, 45727--45747 (2016)\relax
\mciteBstWouldAddEndPuncttrue
\mciteSetBstMidEndSepPunct{\mcitedefaultmidpunct}
{\mcitedefaultendpunct}{\mcitedefaultseppunct}\relax
\EndOfBibitem
\bibitem{doi:10.1063/1.2816914}
Y.~F. Li, B.~Yao, Y.~M. Lu, Z.~P. Wei, Y.~Q. Gai, C.~J. Zheng, Z.~Z. Zhang,
  B.~H. Li, D.~Z. Shen, X.~W. Fan, and Z.~K. Tang, Applied Physics Letters {\bf
  91}(23), 232115 (2007)\relax
\mciteBstWouldAddEndPuncttrue
\mciteSetBstMidEndSepPunct{\mcitedefaultmidpunct}
{\mcitedefaultendpunct}{\mcitedefaultseppunct}\relax
\EndOfBibitem
\bibitem{doi:10.1063/1.3485058}
Y. Li, R. Deng, B. Yao, G. Xing, D. Wang, and T. Wu, Applied Physics Letters {\bf
  97}(10), 102506 (2010)\relax
\mciteBstWouldAddEndPuncttrue
\mciteSetBstMidEndSepPunct{\mcitedefaultmidpunct}
{\mcitedefaultendpunct}{\mcitedefaultseppunct}\relax
\EndOfBibitem
\bibitem{PhysRevB.72.241305}
C.~Morhain, T.~Bretagnon, P.~Lefebvre, X.~Tang, P.~Valvin, T.~Guillet, B.~Gil,
  T.~Taliercio, M.~Teisseire-Doninelli, B.~Vinter, and C.~Deparis, Phys. Rev. B
  {\bf 72}, 241305 (2005)\relax
\mciteBstWouldAddEndPuncttrue
\mciteSetBstMidEndSepPunct{\mcitedefaultmidpunct}
{\mcitedefaultendpunct}{\mcitedefaultseppunct}\relax
\EndOfBibitem
\bibitem{doi:10.1063/1.2962985}
J.~Park, S.~Kim, C.~Kim, S.~Kim, I.~Song, H.~Yin, K.-K. Kim, S.~Lee, K.~Hong,
  J.~Lee, J.~Jung, E.~Lee, K.-W. Kwon, and Y.~Park, Applied Physics Letters
  {\bf 93}(5), 053505 (2008)\relax
\mciteBstWouldAddEndPuncttrue
\mciteSetBstMidEndSepPunct{\mcitedefaultmidpunct}
{\mcitedefaultendpunct}{\mcitedefaultseppunct}\relax
\EndOfBibitem
\bibitem{doi:10.1063/1.2966145}
J.~Park, I.~Song, S.~Kim, S.~Kim, C.~Kim, J.~Lee, H.~Lee, E.~Lee, H.~Yin, K.-K.
  Kim, K.-W. Kwon, and Y.~Park, Applied Physics Letters {\bf 93}(5), 053501
  (2008)\relax
\mciteBstWouldAddEndPuncttrue
\mciteSetBstMidEndSepPunct{\mcitedefaultmidpunct}
{\mcitedefaultendpunct}{\mcitedefaultseppunct}\relax
\EndOfBibitem
\bibitem{5353711}
D.~A. {Mourey}, D.~A. {Zhao}, J.~{Sun}, and T.~N. {Jackson}, IEEE Transactions
  on Electron Devices {\bf 57}(2), 530--534 (2010)\relax
\mciteBstWouldAddEndPuncttrue
\mciteSetBstMidEndSepPunct{\mcitedefaultmidpunct}
{\mcitedefaultendpunct}{\mcitedefaultseppunct}\relax
\EndOfBibitem
\bibitem{doi:10.1889/JSID18.10.753}
D.~A. Mourey, M.~S. Burberry, D.~A. Zhao, Y.~V. Li, S.~F. Nelson, L.~Tutt,
  T.~D. Pawlik, D.~H. Levy, and T.~N. Jackson, Journal of the Society for
  Information Display {\bf 18}(10), 753--761 (2010)\relax
\mciteBstWouldAddEndPuncttrue
\mciteSetBstMidEndSepPunct{\mcitedefaultmidpunct}
{\mcitedefaultendpunct}{\mcitedefaultseppunct}\relax
\EndOfBibitem
\bibitem{Robertson2000}
J.~Robertson, Journal of Vacuum Science \& Technology B: Microelectronics and
  Nanometer Structures {\bf 18} (2000)\relax
\mciteBstWouldAddEndPuncttrue
\mciteSetBstMidEndSepPunct{\mcitedefaultmidpunct}
{\mcitedefaultendpunct}{\mcitedefaultseppunct}\relax
\EndOfBibitem
\bibitem{doi:10.1063/1.124573}
A.~Ohtomo, M.~Kawasaki, I.~Ohkubo, H.~Koinuma, T.~Yasuda, and Y.~Segawa,
  Applied Physics Letters {\bf 75}(7), 980--982 (1999)\relax
\mciteBstWouldAddEndPuncttrue
\mciteSetBstMidEndSepPunct{\mcitedefaultmidpunct}
{\mcitedefaultendpunct}{\mcitedefaultseppunct}\relax
\EndOfBibitem
\bibitem{doi:10.1063/1.1370116}
G.~Coli and K.~K. Bajaj, Applied Physics Letters {\bf 78}(19), 2861--2863
  (2001)\relax
\mciteBstWouldAddEndPuncttrue
\mciteSetBstMidEndSepPunct{\mcitedefaultmidpunct}
{\mcitedefaultendpunct}{\mcitedefaultseppunct}\relax
\EndOfBibitem
\end{mcitethebibliography}
\end{document}